\documentstyle{article}

\begin{document}

\title{Vortex Solutions of a Fermion Maxwell-Chern-Simons Theory%
\footnote{To appear in the proceedings of "International Workshop on Mathematical 
and Physical Aspacts of Nonlinear Field Theories", 1998.}}
\author{Jae Hyung Yee \\ Department of Physics and Institute for Mathematical Sciences\\ 
Yonsei University\\ 
Seoul 120-749, Korea}
\maketitle
\begin{abstract}
We explain how static multi-vortex solutions arise in non-linear field theories,
by taking the non-linear Schr\"odinger equation coupled to Chern-Simons field (Jackiw-Pi model)
and a fermion Chern-Simons theory as simple examples.
We then construct a fermion Maxwell-Chern-Simons theory which has consistent static field equations,
and show that it has the same vortex solutions as the Jackiw-Pi model, but gives rise to 
quite different vortex dynamics.
\end{abstract}

\newpage

\section*{I. Introduction}

Magnetic vortices are found to appear in high temperature superconductors 
as quantized vortex lines in the range of magnetic field between $10^{-2}$ tesla and
$10^2$ tesla \cite{a}. It is known that the behavior of vortices dominates many physical
properties of high temperature superconductors in this range of magnetic field.
These vortices show very interesting dynamical properties:in some range of magnetic 
field and temperature they form vortex lattice or behave as liquid. This phenomenon 
poses an important theoretical challenge in understanding the dynamics of magnetic vortices.

One way to understand the dynamics of vortices is to consider non-linear field theories
that possess classical vortex solutions and to study the dynamics of these solutions 
in the field theoretic framework. There exist many 
field theoretic models that support the static vortex solutions. 
Relativistic models of vortices include the Abelian Higgs model of Nielson and Olesen \cite{b},
the scalar Chern-Simons theory of Hong, Kim and Pac, and Jackiw and Weinberg \cite{c}, 
and the fermion Chern-Simons thoery of Lie and Bhaduri \cite{d}. 
There also exist many non-relativistic models including the Ginzburg-Landau model \cite{e}
which is the non-relativistic limit of the Abelian Higgs model, Jackiw-Pi model \cite{f}
which is the non-relativistic limit of the scalar Chern-Simons theory, 
and non-relativistic spinor Chern-Simons theories \cite{g}.

The most of these models are not soluble in closed form except for the Jackiw-Pi model 
and some of its generalizations. For the study of vortex dynamics in the field theoretic 
framework, however, it would be more convenient to have completely soluble models.
We therefore consider the field theoretic models which are completely soluble.

In the next section we give a brief review of the Jackiw-Pi model to show how to find 
the static self-dual solutions. In section III, the simplest fermion Chern-Simons theory 
which supports the static vortex solutions as solutions of the Liouville equations is presented.
In section IV, we present a fermion Maxwell-Chern-Simons theory which supports static vortex
solutions. We show that although the static solutions are the same as those of Jackiw-Pi model
and the simplest fermion Chern-Simons theory, this model gives rise to quite different 
moduli space dynamics from those simple models. 
We conclude with some discussions in the last section.

\section*{II. Non-linear Schr\"odinger Equation coupled to Chern-Simons Gauge Field}

As a simplest model field theory that possesses vortex solutions we consider the 
non-linear Schr\"odinger field theory coupled to a Chern-Simons gauge field 
described by the Lagrangian \cite{f},
\begin{equation}
 {\cal L} = { \kappa \over 4} \epsilon^{\alpha \beta \gamma } F_{\alpha \beta } A_{\gamma}
          + i \phi^* D_0 \phi - {1 \over 2} {|{\vec{D}} \phi |}^2 + {1 \over {2 \kappa }}(\phi ^* \phi )^2 ,
\label{aaa}
\end{equation}
where  
\begin {eqnarray}
F_{\mu \nu}\!\!\!\! &=& \!\!\!\!\partial_\mu A_\nu - \partial_\nu A_\mu \nonumber \\ 
A_\mu\!\!\!\! &=& \!\!\!\!(A^0 , - \vec{A} ) \nonumber \\
D_0\!\!\!\! &=& \!\!\!\!\partial_0 + i A^0 ,\qquad  \vec{D} = \vec{\nabla} - i \vec{A}. \nonumber
\end {eqnarray} 

One way to see the existence of static solutions is to consider the energy functional 
and to find the field configurations that minimize the energy functional \cite{e}. 
To this end it is convenient to write the Lagrangian (\ref{aaa}) as
\begin{equation}
{\cal L} = -{ \kappa \over 2} \epsilon_{ij} A^i \dot{A}^j
          + i \phi^* \dot{ \phi} -A^0(\phi^* \phi - \kappa F_{12} ) - {1 \over 2} {|{\vec{D}} \phi |}^2 + {1 \over {2 \kappa }}(\phi^* \phi )^2 ,
\label{aab}
\end{equation} 
where indices $i$ and $j$ run for the spatial components 1 and 2. The momentum conjugates 
to the field variables $A^i$ and $\phi$ are defined by 
\begin {eqnarray}
\pi_i \!\!\!\! &=& \!\!\!\! {{\delta {\cal L}} \over {\delta \dot{A}^i }} = {\kappa \over 2} \epsilon_{ij} A^j  \nonumber \\
P \!\!\!\! &=& \!\!\!\! {{\delta {\cal L}} \over {\delta \dot{\phi}}} = i \phi^* ,
\label{aac}
\end {eqnarray} 
respectively. The Lagrangian (\ref{aab}) can then be written as 
\begin{equation}
{\cal L} = \pi_i \dot{A}^i + P \dot{\phi} - {\cal H} - A^0 (\phi^* \phi - \kappa F_{12} ),
\label{aad}
\end{equation} 
where ${\cal H}$ is the Hamiltonian density and the last term represents the Gauss' law constraint,
\begin{equation}
B = \vec{\nabla} \times \vec{A} =  -F_{12} =  -{1 \over \kappa } \phi^* \phi.
\label{aae}
\end{equation} 
The Gauss' law constraint (\ref{aae}) shows that the magnetic field is proportional to the charge density
of the scalar field,
\begin{equation}
\rho = \phi^* \phi.
\label{aaf}
\end{equation} 

From Eqs. (\ref{aab}) and (\ref{aad}) one finds that the Hamiltonian density of the system is given by
\begin{equation}
{\cal H} =  {1 \over 2} |\vec{D} \phi |^2 - {1 \over 2 \kappa} (\phi^* \phi )^2.
\label{aag}
\end{equation} 
Using the identity,
\begin{equation}
|\vec{D} \phi |^2 =  |(D_1 - i D_2 ) \phi |^2 - ( B \rho + \vec{\nabla} \times \vec{J} ),
\label{aah}
\end{equation} 
where $\vec{J}=Im\phi^* \vec{D}\phi$ is the current density, one can write the Hamiltonian density as
\begin{equation}
{\cal H} =  {1 \over 2} |(D_1 - i D_2 ) \phi |^2 - {1 \over 2} {\vec{\nabla} \times \vec{J}}.
\label{aai}
\end{equation} 
The energy of the system then becomes 
\begin{equation}
E = \int d^2 x {\cal H} =  {1 \over 2}  \int d^2 x  \big[ |(D_1 - i D_2 ) \phi |^2  - {\vec{\nabla} \times \vec{J}} \big].
\label{aaj}
\end{equation} 
If the fields are well-behavecd at infinity, then the last term does not contribute to the energy
since it is a surface term. Then the energy of the system is positive definite:
\begin{equation}
E =  {1 \over 2}  \int d^2 x  {|(D_1 - i D_2 ) \phi |}^2  \ge 0,
\label{aak}
\end{equation} 
where the equality is called the Bogomol'nyi bound. Note that the coupling constant of the quartic 
interaction of the scalar field is so chosen that the energy functional is positive definite.

Eq.(\ref{aak}) shows that the minimum energy configurations of the system 
are determined by the static first-order differential equation,

\begin{equation}
(D_1 - i D_2 ) \phi =  0,
\label{aal}
\end{equation} 
which is called the self-dual equation.

To find the static soliton solutions one has to solve the self-dual equation (\ref{aal}) together with
the Gauss' law constraint (\ref{aae}). To do this we write the scalar field as 
\begin{equation}
\phi =  \rho^{1 \over 2} e^{i\omega}.
\label{aam}
\end{equation} 
Substituting (\ref{aam}) into (\ref{aal}) one finds
\begin{equation}
A_- =  - i {\phi}^{-1} \partial_- \phi =  -{i \over 2} \partial_- \ln {\rho} + \partial_- \omega ,
\label{aan}
\end{equation} 
or 
\begin{equation}
\vec{A} =  \vec{\nabla} \omega - {1 \over 2 } {\vec{\nabla} \times \ln{\rho} },
\label{aao}
\end{equation} 
where $A_{\pm} = A_1 \pm i A_2 $ and $ \partial_{\pm} = \partial_1 \pm i \partial_2 $. 
This shows that the gauge field $\vec{A}$ is completely determined by the scalar field. 
From (\ref{aan}) we obtain
\begin{equation}
\partial_- A_+ -  \partial_+ A_- =  2i \vec{\nabla} \times {\vec{A}} =  2i B =  i \vec{\nabla}^2 \ln \rho ,
\label{aap}
\end{equation} 
which, upon substituting into the Gauss' law constraint, reduces to the Liouville equation,
\begin{equation}
\nabla^2 \ln{\rho} =  - {2 \over \kappa} \rho,\quad \kappa > 0,
\label{aaq}
\end{equation} 
which is completely integrable.

The simplest solution of the Liouville equation (\ref{aaq}) is the spherically symmetric solution,
\begin{equation}
\rho (\vec{r}) =  {{4 \kappa {N^2}} \over r^2 } {\big[{\big({{r_0} \over r }\big)}^N + {\big({r \over {r_0}} \big)}^N \big]}^{-2},
\label{aar}
\end{equation} 
where $r_0$ and $N$ are the constants representing the scale and the flux number of the soliton, respectively.
To find the restriction on the number $N$, we observe that the regularity of the solution at the origin 
and at infinity requires $N \ge 1$. And the single-valuedness of the scalar field $\phi$
requires $N$ to be an integer. Note that the regularity of the gauge field at the origin,
\begin{equation}
A_i (\vec{r})  \stackrel{r \to 0}{\longrightarrow} \partial_i \omega - {1 \over 2} \epsilon_{ij} {r^j \over r^2 } (N-1),                 
\label{aas}
\end{equation} 
determines the function $\omega$:
\begin{equation}
\omega =  (1 - N) \theta.
\label{aat}
\end{equation}
We thus find the static solution,
\begin{equation}
\phi (\vec{r}) =  {{2 \sqrt{\kappa} N} \over r} {\big[{\big({{r_0} \over r} \big)}^N + {\big({r \over {r_0}} \big)}^N \big]}^{-1} e^{i(1-N) \theta },
\label{aau}
\end{equation} 
with $\vec{A}$ determined by Eq.(\ref{aao})

From the explicit solution (\ref{aau}) one finds the magnetic flux of the soliton solution,
\begin{equation}
\Phi =   \int d^2 x B =  - {1 \over \kappa} \int d^2 x \rho =  - 2 \pi ( 2N ),
\label{aav}
\end{equation} 
the electric charge,
\begin{equation}
Q =  - \kappa \Phi ,
\label{aaw}
\end{equation} 
and the angular momentum,
\begin{equation}
J =  2 \pi \kappa ( 2N ).
\label{aax}
\end{equation} 
This shows that the solution (\ref{aar}) and (\ref{aau}) represents the charged vortex solution 
with quantized magnetic flux.

The solution (\ref{aar}) represents the case of $N$ solitons superimposed at the origin. 
The general solution of the Liouville equation is also known:
\begin{equation}
\rho (\vec{r}) =  {{4\kappa {|f^{'} (z)|}^2 } \over {{(1 + {|f(z)|}^2 )}^2 }} ,
\label{aay}
\end{equation} 
where $z = r e^{i \theta } $ and $ f(z) $ is an arbitrary function of $z$ such that $\rho $ is non-singular.
$f(z)$ for the most general solution can be written as 
\begin{equation}
f(z) =  \sum_{i=1}^{N} { c_i \over {z-z_i }} ,
\label{aaz}
\end{equation} 
where $c_i $ and $z_i $ are arbitrary constants. This solution contains $4N$ arbitrary parameters which
represent $2N$ position parameters, $N$ scale parameters and $N$ phase variables of the solitons.
In fact, index theorem confirms that $(4N)$ is the maximal number of parameters contained in the general 
solution \cite{h} .

The Jackiw-Pi model described above is invariant under the Galilean transformation, 
and is the non-relativistic limit of the relativistic scalar Chern-Simons theory \cite{c} ,
\begin{equation}
 {\cal L}_{rel}  =  { \kappa \over 4} \epsilon^{\alpha \beta \gamma } F_{\alpha \beta } A_{\gamma}
          + D_{\mu} \phi D^{\mu} \phi^* - {1 \over {\kappa }^2 } |\phi|^2 (|\phi|^2 - \kappa )^2 .
\label{aba}
\end{equation} 
This theory possesses two types of soliton solutions:
the topological solitons in the symmetry broken sector, and the non-topological solitons in the symmetric sector.
The magnetic flux of the topological solitons is quantized as 
\begin{equation}
\Phi =  2 \pi (N-1),
\label{abb}
\end{equation} 
while the magnetic flux of the non-topological solitons is not quantized:
\begin{equation}
\Phi =  2 \pi (N + \alpha ),
\label{abc}
\end{equation} 
where $\alpha $ is an arbitrary parameter such that $ \alpha \ge N $.

The non-relativistic limit of the model (\ref{aba}) in the symmetric sector reduces to the Jackiw-Pi model.
Why then the non-relativistic solitons have quantized flux while magnetic flux of the relativistic case 
is not quantized? The answer lies in the fact that the Jackiw-Pi model has inversion symmetry which is not
respected by the relativistic model \cite{i}.  
To see this note that the charge density satisfies the equation,
\begin{equation}
{{\nabla}^2} {\ln{\rho}} =  \left\{ 
\begin{array}{ll}  
{4m^2 \over \kappa } \rho {( 1 - {\rho \over \kappa})}, & \mbox{relativistic case}  \cr
  - {2 \over \kappa } \rho , & \mbox{non-relativistic case.}  
\end{array} \right.
\label{abd}
\end{equation} 
One can easily show that the Liouville equation is invariant under the inversion transformation:
\begin{eqnarray}
r \!\!\!\! &\to& \!\!\!\! {1 \over r} ,  \qquad \theta \to \theta \nonumber \\
\rho (r) \!\!\!\! &\to& \!\!\!\! \rho \big({1 \over r }\big) =  r^4 \rho (r),
\label{abe}
\end{eqnarray} 
while the relativistic equation is not. The charge density for both models behaves as 
\begin{equation}
\rho (r) \longrightarrow \left\{ 
\begin{array}{ll} 
r^{2(N-1)},  & \mbox{as} \quad r \to 0 \\
r^{-2(\alpha + 1) }, & \mbox{as} \quad r \to \infty ,
\end{array} \right.
\label{abf}
\end{equation} 
and thus the magnetic flux is given by 
\begin {eqnarray}
- \Phi \!\!\!\! &=& \!\!\!\!  \int_{r \to \infty } \vec{A} \cdot d\vec{l} \nonumber \\
      \!\!\!\! &=& \!\!\!\!  \int_{r \to \infty } (- \vec{\nabla} \omega + {1 \over 2}  \vec{\nabla} \times  \ln \rho ) \cdot dl \\
      \!\!\!\! &=& \!\!\!\!  2 \pi (n-1) + 2 \pi ( \alpha + 1 ). \nonumber
\label{abg}
\end {eqnarray} 
Due to the inversion symmetry (\ref{abe}) of the Jackiw-Pi model, the behavior of $\rho (\vec{r})$ at the
origin and at infinity (\ref{abf}) must be related. This give rises to the relation $ \alpha = N$, and 
explains the quantization of flux numbers in the case of the non-relativistic theory.

\section*{III. Vortex Solutions in a Fermionic Chern-Simons Theory}

Vortex phenomena in realistic systems are known to be dominated by the electronic structure of the systems. 
Thus in attempting to understand the vortex dynamics it would be preferrable to use the fermionic 
field theories that support the static vortex solutions. The simplest fermionic Chern-Simons theory 
that supports the vortex solutions is the one proposed by Li and Bhaduri \cite{d}, 
described by the Lagrangian,
\begin{equation}
{\cal L} =  { \kappa \over 4} \epsilon^{\mu \nu \alpha } F_{\mu \nu } A_{\alpha}
          + i \bar{\psi} {\gamma}^{\mu} {\partial}_{\mu} \psi - m \bar{\psi} \psi + e A_{\mu} j^{\mu} ,
\label{abh}
\end{equation} 
where $\psi$ is a two-component spinor field, the Dirac matrices are chosen to be 
\begin{equation}
{\gamma}^0 =  {\sigma}^3 ,  {\gamma}^1 =  i{\sigma}^1 ,  {\gamma}^2 =  i{\sigma}^2 
\label{abi}
\end{equation} 
in terms of the Pauli matrices ${\sigma}^i $ , and 
\begin{equation}
j^{\mu} =   \bar{\psi} {\gamma}^{\mu} \psi , \qquad  \bar{\psi} =  {\psi}^{\dagger} {\gamma}^0 .  
\label{abj}
\end{equation} 
We will denote the charge density as $j^0 =  {\psi}^{\dagger} {\psi} = \rho $.

To find the static vortex solutions in this system, it is simpler to start from the field equations,
\begin{eqnarray}
  {\gamma}^{\mu} ( i {\partial}_{\mu} + e A_{\mu} ) \psi - m \psi = 0 \!\!\!\!&,& \nonumber \\
  { \kappa \over 2} \epsilon^{\mu \nu \alpha } F_{\nu \alpha } = -e j^{\mu} \!\!\!\!&.&
\label{abk}
\end{eqnarray}
The second equation of (\ref{abk}) can be decomposed into the Gauss' law constraint,
\begin{equation}
B =  F_{12} =  {e \over \kappa } \rho , 
\label{abl}
\end{equation} 
and 
\begin{equation}
E^i =  F_{0i} =   {e \over \kappa } {\epsilon}^{ij} j^j .
\label{abm}
\end{equation} 

By writing 
\begin{equation}
\psi =  \left( 
\begin{array}{c}  {{\psi}_+ ( \vec{x},t)} \\ {{\psi}_- ( \vec{x},t)} 
\end{array} \right),
\label{abn}
\end{equation} 
the fermion field equation, the first of (\ref{abk}), reduces to the two coupled equations for
${\psi}_+ ( \vec{x},t)$ and ${\psi}_- ( \vec{x},t)$ :
\begin {eqnarray}
(i {\partial}_0 - m ) {\psi}_+ ( \vec{x},t) \!\!\!\! &=& \!\!\!\! (D_1 - i D_2 ) {\psi}_- ( \vec{x},t)  \nonumber \\
-(i {\partial}_0 + m ) {\psi}_- ( \vec{x},t) \!\!\!\! &=& \!\!\!\! (D_1 + i D_2 ) {\psi}_+ ( \vec{x},t), 
\label{abo}
\end {eqnarray} 
where $D_i = {\partial}_i - ieA^i$ .

We now seek the stationary solution of the  form,
\begin{equation}
\psi =  {\psi}_+ ( \vec{x}) \left(
\begin{array}{c} 1 \\ 0 \end{array}\right) e^{-i E_f t } ,
\label{abp}
\end{equation} 
where $E_f$ is a constant. Then the fermion field equation reduces to 
\begin{equation}
(E_f - m) {\psi}_+ ( \vec{x}) =  0,
\label{abq}
\end{equation} 
which determines the constant $E_f = m$, and the self-dual equation,
\begin{equation}
(D_1 + i D_2 ) {\psi}_+ ( \vec{x})  =  0.
\label{abr}
\end{equation} 
Note that this is the same type of self-dual equation as that of the Jackiw-Pi model.

If we take ,
\begin{equation}
\psi =  {\psi}_- ( \vec{x}) \left(
\begin{array}{c} 0 \\ 1 \end{array}\right) e^{-i E_f t } ,
\label{abs}
\end{equation} 
instead, the fermion field equation reduces to 
\begin{equation}
(E_f + m) {\psi}_- ( \vec{x}) =  0
\label{abt}
\end{equation} 
which determines $E_f = -m$ and the self-dual equation,
\begin{equation}
(D_1 - iD_2) {\psi}_- ( \vec{x}) =  0.
\label{abu}
\end{equation} 

To find the static solutions we choose the gauge, $A^{0}=0$, and take $\vec{A}$ to be static. 
Then the both sides of Eq.(\ref{abm}) consistently vanish, 
and the spatial components of matter current vanish for spinor fields (\ref{abp}) and (\ref{abs}), 
since ${\gamma}^i$'s are off-diagonal. Thus for the upper component spinor field (\ref{abp}), 
the field equations reduce to the self-dual equation(\ref{abr}) and the Gauss' law constraint (\ref{abl}).
If we write the spinor field as 
\begin{equation}
{\psi}_+ ( \vec{x}) = {\rho}^{1 \over 2} e^{i\omega} ,
\label{abv}
\end{equation} 
Eq.(\ref{abr}) reduces to 
\begin{equation}
eA_+ =  -{i \over 2} {\partial}_+ \ln{\rho} +  {\partial}_+ \omega ,
\label{abw}
\end{equation} 
which determines the gauge field in terms of the matter fields:
\begin{equation}
\vec{A} =  \vec{\nabla} \omega - {1 \over 2}  \vec{\nabla} \times \ln \rho .
\label{abx}
\end{equation} 
From Eq.(\ref{abw}) together with the Gauss' law constraint (\ref{abl}), one finds that the matter 
charge density satisfies the Liouville equation:
\begin{equation}
{\nabla}^2 \ln{\rho} =  - {2e^2 \over \kappa} \rho .
\label{aby}
\end{equation} 
Thus if we take the upper-component for spinor field (\ref{abp}), $\kappa > 0$ is required in order to 
have non-singular positive charge density $\rho$.

If we take the lower-component for the matter field, Eq(\ref{abs}), on the other hand, the field equations
reduce to 
\begin{equation}
{\nabla}^2 \ln \rho =  {2e^2 \over \kappa} \rho ,
\label{abz}
\end{equation} 
where $\kappa$ must be negative for the regularity of the charge density $\rho$.

Since the matter charge density $\rho$ satisfies the same Liouville equation (\ref{aby}) and (\ref{abz})
as that of Jackiw-Pi model, we find the same structure of the static vortex solutions except that the
energy of the solution is now given by 
\begin{equation}
E = \pm m \int d^2 x {\rho}_{\pm},
\label{aca}
\end{equation} 
where ${\rho}_{\pm} = {{\psi}_{\pm}}^{\dagger}(\vec{x}) {{\psi}_{\pm}}(\vec{x})$ for the upper and lower 
component matter fields, respectively. We thus find that the fermion Chern-Simons theory (\ref{abh}) supports
the finite energy static vortex solutions with quantized magnetic flux and charge.

In general field theories with Chern-Simons term, parity is known to be violated by the Chern-Simons term.
One can restore the parity invariance by introducing appropriate parity partners for each field 
in the theory. A parity invariant fermion Chern-Simons theory is proposed by Hagen \cite{j}.
This parity invariant fermion Chern-Simons theory also supports the static vortex solutions 
as solutions of Liouville equation \cite{k}.

\section*{IV. Vortex Solutions of a Fermion Maxwell-Chern-Simons Theory}

To find a field theoretic model which supports the vortex solutions with physically more interesting
properties, we introduce a Maxwell term to the gauge field part of the Lagrangian  and a couple of 
new interaction terms \cite{l}:
\begin{equation}
 {\cal L} =  -{1 \over 4} F^{\mu \nu} F_{\mu \nu} + { \kappa \over 4} \epsilon^{\alpha \beta \gamma } F_{\alpha \beta } A_{\gamma}
          + i \bar{\psi} {\gamma}^{\mu} {\partial}_{\mu} \psi - m \bar{\psi} \psi + eA_{\mu} (J^{\mu} + lG^{\mu} ) + {1 \over 2} g (\bar{\psi} \psi)^2,
\label{acb}
\end{equation} 
where $F_{\mu \nu} = {\partial}_{\mu} A_{\nu} - {\partial}_{\nu} A_{\mu}, J^{\mu} = \bar{\psi} {\gamma}^{\mu} \psi $
 and the new current $G^{\mu}$ is defined by 
\begin{equation}
G^{\mu} =  \epsilon^{\mu \nu \rho} {\partial}_{\nu} J_{\rho}.
\label{acc}
\end{equation} 
The new gauge coupling term can be written as 
\begin {eqnarray}
A_{\mu}G^{\mu} \!\!\!\! &=& \!\!\!\! {\partial}_{\mu} A_{\nu} \epsilon^{\mu \nu \rho} J_{\rho} + \mbox{surface term} \nonumber \\
               \!\!\!\! &=& \!\!\!\! F_{\mu \nu} \epsilon^{\mu \nu \rho} J_{\rho} + \mbox{surface term}, 
\label{acd}
\end {eqnarray} 
which is the magnetic moment coupling in 3-dimensional space-time. This is the reason why this new 
coupling is called an anomalous magnetic moment coupling.

If the gauge field part of the Lagrangian (\ref{acb}) is eliminated, then the theory reduces to the 
3-dimensional Gross-Neveu model. If we let $l \to 0, g \to 0$ and $e^2 \to \infty$ with 
${\kappa \over e^2}$ fixed, the theory becomes the fermion Chern-Simons theory discussed in the last section.

Field equations of the new theory are 
\begin{eqnarray}
& {\partial}_{\nu} F^{\mu \nu} + { \kappa \over 2}  \epsilon^{\mu \nu \rho} F_{\nu \rho} =  -e(J^{\mu} + lG^{\mu}) \nonumber \\
& {\gamma}^{\mu} ( i {\partial}_{\mu} + e A_{\mu} ) \psi - m \psi +  g (\bar{\psi} \psi) \psi - el\epsilon^{\mu \nu \rho} ({\partial}_{\nu} A_{\mu}) {\gamma}_{\rho} \psi =  0,
\label{ace}
\end{eqnarray} 
the first of which is the field equation for a massive gauge field in 3-dimensions.

To find static solutions we choose the gauge, $A^0 = 0$, and take $\vec{A}$ to be static. 
As in the last section we write the spinor field in component form,  
\begin{equation}
\psi = \left( 
\begin{array}{c} {\psi}_+ (\vec{x}) \\ {\psi}_- (\vec{x}) \end{array} \right) e^{-i E_f t} 
\label{acf}
\end{equation} 
where $E_f$ is a constant. Then the spinor field equation reduces to the coupled equations,
\begin{eqnarray}
\big[E_f - m + g( |{\psi}_+|^2 - |{\psi}_-|^2 ) - el{\epsilon}^{ij} {\partial}_j A_i \big] {\psi}_+  \!\!\!\! &=& \!\!\!\!  D_- {\psi_-}  \nonumber \\
\big[-E_f - m + g( |{\psi}_+|^2 - |{\psi}_-|^2 ) + el{\epsilon}^{ij} {\partial}_j A_i \big] {\psi}_-  \!\!\!\! &=& \!\!\!\!  D_+ {\psi_+}, 
\label{acg}
\end{eqnarray} 
where $D_{\pm} = D_1 \pm iD_2$, and $D_i = {\partial}_i - ieA^i$.

If we take the upper-component spinor field,
\begin{equation}
\psi =  {\psi}_+ ( \vec{x}) \left(
\begin{array}{c} 1 \\ 0 \end{array}
\right) e^{-iE_f t } ,
\label{ach}
\end{equation} 
then Eq.(\ref{acg}) reduces to 
\begin {eqnarray}
[E_f - m + g{\rho}_+ - el{\epsilon}^{ij} {\partial}_j A_i ] {\psi}_+ \!\!\!\! &=& \!\!\!\! 0  \nonumber \\
D_+ {\psi}_+ \!\!\!\! &=& \!\!\!\! 0,
\label{aci}
\end {eqnarray} 
where ${\rho}_+ = {J_+}^0 = |{\psi}_+|^2 $ is the charge density of the matter field. For this choice of the 
spinor field, we find,
\begin {eqnarray}
{J_+}^i \!\!\!\! &=& \!\!\!\!  {\bar{\psi}}_+ {\gamma}^i {\psi_+} =  0 \nonumber \\
G^0 \!\!\!\! &=& \!\!\!\! 0  \\
G^i \!\!\!\! &=& \!\!\!\! {\epsilon}^{ij} {\partial}_j {\rho}_+ , \nonumber
\label{acj}
\end {eqnarray} 
since ${\gamma}^i$'s are off-diagonal matrices. Then the gauge field equation reduces to 
\begin {eqnarray}
{\epsilon}^{ij} F_{ij} \!\!\!\! &=& \!\!\!\! -{2e \over \kappa} {\rho}_+ \nonumber \\
{\partial}_j F^{ij} \!\!\!\! &=& \!\!\!\! elG^i =  el{\epsilon}^{ij} {\partial}_j {\rho}_+ .
\label{ack}
\end {eqnarray} 
For these two equations to be consistent, the coupling constants $l$ and $\kappa$ must be related by
\begin{equation}
l = -{1 \over \kappa}.
\label{acl}
\end{equation} 
This shows that, for the fermion Maxwell-Chern-Simons theory (\ref{acb}) to have consistent static field equations, 
one need to introduce the anomalous magnetic moment coupling term.

To find static solutions, therefore, one has to solve Eqs. (\ref{aci}) and (\ref{ack}), 
which now reduce to 
\begin{eqnarray}
(E_f - m + g{\rho}_+ + {e \over \kappa}{\epsilon}^{ij} {\partial}_j A_i ] {\psi}_+ \!\!\!\! &=& \!\!\!\!  0 \nonumber \\
D_+ {\psi}_+ \!\!\!\! &=& \!\!\!\!  0  \\
B =  - F_{12} \!\!\!\! &=& \!\!\!\!  {e \over \kappa} \rho .\nonumber
\label{acm}
\end{eqnarray} 
Note that, for the first equation of Eq.(\ref{acm}) to be consistent with the other two equations, 
one need to require,
\begin {eqnarray}
E_f = m \nonumber \\
g = -{e^2 \over {\kappa}^2 } .
\label{acn}
\end {eqnarray}
In other words, in order for the theory (\ref{acb}) to have consistent static self-dual solutions, 
the quartic coupling constant of the matter field must be determined by the Chern-Simons 
coupling by (\ref{acn}) as in the case of the Jackiw-Pi model.

To solve the second equation of (\ref{acm}), we write
\begin{equation}
{\psi}_+ =  {{\rho}_+}^{1 \over 2} e^{i\omega}.
\label{aco}
\end{equation}
Then this equation determines the gauge field in terms of the matter field:
\begin {eqnarray}
eA_+ \!\!\!\! &=& \!\!\!\! -{i \over 2} {{\rho}^{-1}_+} {\partial}_+ {{\rho}_+} + {\partial}_+ \omega \nonumber \\
eA_- \!\!\!\! &=& \!\!\!\! {i \over 2} {{\rho}^{-1}_+} {\partial}_- {{\rho}_+} + {\partial}_- \omega,
\label{acp}
\end {eqnarray}
from which one finds,
\begin{equation}
{\partial}_- A_+ - {\partial}_+ A_- =  2iB =  -{i \over e}{\partial}_+{\partial}_- \ln{{\rho}_+}.
\label{acq}
\end{equation}
By substituting this equation to the Gauss' law constraint, the third equation of (\ref{acm}),
one finds the Liouville equation for the matter charge density: 
\begin{equation}
{\nabla}^2 \ln{{\rho}_+} = - {2e^2 \over \kappa} {\rho}_+ ,
\label{acr}
\end{equation}
where $\kappa$ must be positive for the regularity of the matter density. This shows that the fermion
Maxwell-Chern-Simons theory (\ref{acb}) also supports the static vortex solutions as solutions of Liouville equation.

If we choose the lower component spinor field,
\begin{equation}
\psi = {\psi}_- ( \vec{x}) \left(
\begin{array}{c} 0 \\ 1 \end{array}
\right) e^{-i E_f t } ,
\label{acs}
\end{equation} 
the field equations reduce to 
\begin{eqnarray}
(E_f + m + g{\rho}_- + {e \over \kappa} {\epsilon}^{ij} {\partial}_j A_i ) {\psi}_- = 0 \nonumber \\
D_- {\psi}_- = 0 \\
B = {e \over \kappa} {\rho}_- , \nonumber 
\label{act}
\end{eqnarray} 
where ${\rho}_- = {{\psi}^{\dagger}}_- {\psi}_- $. 
For these equations to be consistent, we also need to require,
\begin {eqnarray}
E_f = -m \nonumber \\
g = -{e^2 \over {\kappa}^2 }.
\label{acu}
\end {eqnarray}
By writing ${\psi}_- = {{\rho}_-}^{1 \over 2} e^{i\omega}$, we again obtain the Liouville equation
for the matter density,
\begin{equation}
{\nabla}^2 \ln{{\rho}_-} =  - {2e^2 \over \kappa} {\rho}_- ,
\label{acv}
\end{equation}
where $\kappa < 0$ is required for the regularity of the density ${\rho}_- $.

We now consider the energy of the static vortex solutions. The Hamiltonian density is given by
\begin{equation}
{\cal H} =  {\pm} m {\rho}_{\pm} + {1 \over 2} {F_{12}}^2 - {1 \over 2} g{{\rho}_{\pm}}^2 + {{e^2 l } \over \kappa} {{\rho}_{\pm}}^2 .
\label{acw}
\end{equation}
Due to the consistency requirements (\ref{acl}), (\ref{acn}), and (\ref{acu}), the last three terms of (\ref{acw}) cancel out,
and the energy of the static solutions reduces to 
\begin{equation}
E =  \pm m \int d^2 {{\rho}_{\pm}} =  {{2\kappa} \over e^2 } m N_{\pm} , 
\label{acx}
\end{equation}
where $2N_{\pm}$ denotes the flux number of the solutions.
The cancellation of the quartic terms in the Hamiltonian density is also reminiscent of that in the Jackiw-Pi model.

One may wonder why one studies such a complicated model with an unusual interaction term if it gives the 
same static vortex solutions as those of the Jackiw-Pi model or the simple fermion Chern-Simons theory.
Although the theory (\ref{acb}) gives the same static solutions as those of the Jackiw-Pi model, 
the moduli space dynamics of this model is quite different from the Jackiw-Pi model because 
of the Maxwell term in the Lagrangian. This may give more interesting, and hopefully more realistic vortex dynamics.

The theory we have described is a $U(1)$ gauge theory. This theory can be generalized to $SU(N)$ gauge 
theories as has been done by Jackiw and Pi \cite{f} for the  Jackiw-Pi model. 
There exist many possibilities to formulate corresponding $SU(N)$ gauge theories as a generalization of the fermion
Maxwell-Chern-Simons theory. The simplest example is to take adjoint representation for the matter field and use the 
ansatz that the matter density and the gauge field $A_{\mu}$ lie in the Cartan subalgebra of the grhoup \cite{m}.
One then finds the equations satisfied by the components of matter density,
\begin{equation}
{\nabla}^2 \ln{{\rho}_\alpha} =  \mp {2e \over \kappa} \sum_{\beta = 1}^{r} K_{\alpha\beta} {\rho}_{\beta} ,
\label{acy}
\end{equation}
where $K_{\alpha\beta} $ is the Cartan matrix. Eq.(\ref{acy}) is not completely integrable in general, 
but can be integrated numerically \cite{f}. For the $SU(2)$ case, Eq.(\ref{acy}) reduces to the Liouville
equation, and it corresponds to an embeding of $U(1)$ in $SU(2)$.

\section*{V. Discussions}

We have constructed a fermion Maxwell-Chern-Simons theory that possesses completely integrable multi-vortex solutions,
and have shown that the solutions are the same as those of Jackiw-Pi model. 
Although the model gives the same static solutions as the Jackiw-Pi model, the moduli space dynamics of the static 
solutions will be quite different due to the Maxwell term in the Lagrangian. 
The Maxwell term is quadratic in the time-derivatives of gauge field, i.e., ${1 \over 2} \dot{A_i} \dot{A_i}$, 
which gives rise to a non-trivial contribution to the moduli space metric, while the Chern-Simons term is linear 
in the time-derivatives of gauge field.
This may give an interesting vortex dynamics and may be more relevant in understanding the vortex phenomena 
arising in the high temperature superconductors.

Similar construction can be done for more relatistic, non-relativistic fermion gauge theories.
Some models have already been constructed and their static solutions have been studied \cite{g}.
These models provide simple starting point in studying more realistic vortex dynamics in the field theoretic framework.

The most well-known method of studying vortex dynamics is that of moduli space dynamics for slowly moving 
vortices proposed by Manton \cite{n}. This method, however, has a difficulty in obtaining the moduli 
space metric for the theories with kinetic terms linear in time-derivatives, such as Chern-Simons theory, 
relativistic fermion theories and general non-relativistic field theories.
This difficulty can be avoided by using the method proposed by Bak and Lee \cite{o}, where one integrates 
out the momentum variables to make the kinetic terms quadratic in time-derivatives of field variables.
One can then use the Manton's method to obtain the correct moduli space metric.

Another way to study the vortex dynamics is the so-called string formulation of vortex dynamics \cite{p}.
The authors of ref.\cite{p} note that at the centers of vortices the field function(scalar field $\phi$ 
in the Jackiw-Pi model and spinor field $\psi$ in the fermionic models) vanishes, and treat the points 
of zeros in 3-dimensional space-time (or string of zeros in 4-dimensional space-time) as fundamental objects.
They then write exact equations of motion for points(or strings) in terms of fields that surround them. 
These equations of motions are rather complicated and need approximation for practical applications.

As mentioned above, it is not a simple matter to use these methods to study vortex dynamics in the field 
theoretic framework. The field theoretic models discussed here have static vortex solutions in closed form, 
and it is relatively simple to apply these methods to understand the dynamics of magnetic vortices.

\vspace{2.5cm}
\section*{Acknowledgements}
\vspace{0.5cm}
The author would like to thank Professor Dongho Chae, the organizer of the Workshop.
This work was supported in part by the Korea Science and Engineering Foundation, under Grant
No. 97-07-02-02-01-3 and 065-0200-001-2, the Center for Theoretical Physics(SNU),
and the Basic Science Research Institute Program, Ministry of Education, 
under project No. BSRI-97-2425.

\end{document}